\documentclass[12pt]{article}
\usepackage{fullpage,authblk,cite}
\usepackage{amsmath,amsthm,graphicx,amssymb,verbatim,listings}
\usepackage{wasysym,color,enumitem,mathcomp}
\usepackage[T1]{fontenc}     
\usepackage{lmodern, tipa}   
\usepackage[ pdftex, plainpages = false, pdfpagelabels,
                 pdfpagelayout = useoutlines,
                 bookmarks,
                 bookmarksopen = true,
                 bookmarksnumbered = true,
                 breaklinks = true,
                 linktocpage=all,
                 pagebackref=false,
                 colorlinks = true,
                 linkcolor = Blue,
                 urlcolor  = blue,
                 citecolor = BrickRed,
                 anchorcolor = green,
                 hyperindex = true,
                 hyperfigures
                 ]{hyperref}
\usepackage[usenames, dvipsnames]{xcolor}
\usepackage{xifthen} 

\newcommand{\arxiv}[2][]{\ifthenelse{\isempty{#1}}{\href{http://arxiv.org/abs/#2}{{\tt arXiv:\allowbreak{}#2}}} {\href{http://arxiv.org/abs/#2}{{\tt arXiv:\allowbreak{}#2 [#1]}}}}

\newcommand{\booktitle}{\textsl}
\newcommand{\hrefdoi}[2]{\href{https://dx.doi.org/#1}{#2}}

\begin{document}
\title{Misreading EPR: Variations on an Incorrect Theme}
\author[$\dag$]{Blake C.\ Stacey}
\affil[$\dag$]{Physics Department, University of
    Massachusetts Boston\protect\\ 100 Morrissey Boulevard, Boston MA 02125, USA}

\date{\today}

\maketitle

\begin{abstract}
  Notwithstanding its great influence in modern physics, the EPR
  thought-experiment has been explained incorrectly a surprising
  number of times.
  \end{abstract}

\section{Introduction}

Expositions of the history of physics tend unhappily often to
degenerate into caricature.  This can happen because we are trying to
teach students who don't yet have the background knowledge to
understand how discoveries were actually made~\cite[\S
  11.1]{Stacey:2015}. For example, the physicists who pioneered
quantum mechanics knew more classical physics than modern
undergraduates do, so it is really not possible to lay out the details
of what those pioneers did and why.  The problem is compounded when
textbooks pass along these oversimplifications.  We physicists are
generally honest enough that we don't deliberately fabricate history
outright, but we are not trained to be historians.  Getting to the
point where we can assign homework is almost always a higher priority
than properly chronicling the development of our field.  And the
trouble is compounded yet again when our subject is popularized, a
process which often appears to lack any quality control whatsoever.

All this is said as explanation, not excuse.

Speaking only for myself, I thought I had gotten used to this
situation.  But then I started turning up examples, one after the
other, of people getting the EPR experiment wrong.  The
thought-experiment of Einstein, Podolsky and Rosen --- one of the most
significant and influential conceptions of twentieth-century physics
--- fundamentally mangled!  The issue with these passages was not
their \emph{interpretation} of EPR, an area where we could debate
endlessly, with or without the jazz cigarettes.  No, what they were
getting wrong was the plain statement of what EPR themselves had said.

Some of the examples we will see come from popularizations of science.
Accordingly, in Section~\ref{sec:newyork} we will go through the EPR
thought-experiment in the manner of a glossy pop-science magazine of
yore.  All of the errors we will put on display can be appreciated
with that background.  Bits of mathematical notation in the more
``serious'' exhibits may make their prose harder to follow, but the
formulae are only a smokescreen over the conceptual problems.

After establishing what the EPR thought-experiment is, in Sections
\ref{sec:becker} through \ref{sec:popper} we will explore what it
isn't.  Each section will treat a published explanation of EPR that
gets it wrong.  To the best of my knowledge, these instances could
well be independent in origin.  Their sources run the gamut from
respected popularizations to a physics textbook.

The EPR thought-experiment was a critique of a philosophical view
promoted by Bohr.  In Section~\ref{sec:interlude}, we will take a
brief look at Bohr's reply, which is known for its obscurity ---
rightly so, I will argue.  It can be substantially improved upon, even
while staying within the Bohrian tradition of thought (if not
necessarily of prose). We will conclude with a discussion of the
limitations of the EPR thought-experiment, and how moving beyond it
motivates a modern view of quantum theory that owes a debt both to
Einstein and to Bohr while pledging fealty to neither.

\section{EPR, Very Quickly}
\label{sec:newyork}
The thought-experiment of Einstein, Podolsky and
Rosen~\cite{Einstein:1935} was a response to a philosophical viewpoint
on quantum mechanics chiefly associated with Werner Heisenberg and
Niels Bohr.  We might summarize that view with the following slogan:
``We can't say that a quantum system has a property --- like a
position, speed or amount of energy --- unless we set up the lab
equipment to measure that property.''  While Heisenberg and Bohr
differed philosophically in significant ways~\cite{Camilleri:2015}, it
is fair to call this a common theme between them, and this is the idea
which EPR (as they are ubiquitously known) wanted to critique. Their
goal was to present a scenario in which a quantum system can be
demonstrated to have a property \emph{whether or not} that property is
measured; thus, while quantum theory is correct in that the math works
in practice, the Bohr--Heisenberg story cannot be the right way to
understand it. In EPR's own terminology, their conclusion is that
quantum theory must be ``incomplete''.

The EPR thought-experiment rests upon the fact that in quantum
physics, it is possible to prepare a pair of objects in such a way
that, once we perform a measurement of our choice on one object, we
can then make a completely confident prediction about the outcome of a
corresponding measurement on the other.  We make a choice and take
action here, and then we can foretell what will happen when we act
over there.  The original EPR presentation considered position and
momentum as the possible ``observables'': Of each particle, we can ask
where it is, or alternatively, how it is moving. These are a pair of
``observables'' to which the \emph{uncertainty principle}
applies. According to the rules of quantum mechanics, there is a
tradeoff of predictability between the two. For any single quantum
particle, if we can make a precise prediction of a position
measurement, we cannot make a precise prediction of a momentum
measurement, and vice versa. This raises the question of how to
\emph{interpret} that tradeoff. Is it just that we are big and
electrons are small, so our attempts to measure them are always clumsy
and disrupt what they're doing\ldots\ Or is the truth something
stranger?  Does a quantum particle \emph{have} a momentum before we go
and measure it?

We prepare two particles following the EPR scheme, keep one of them
close at hand and send the other far away --- let's say, in the
direction of Mars.  Then we choose which question to ask of the
particle that we kept nearby: \emph{Where exactly are you?}\ or
\emph{How exactly are you moving?} Suppose that we do the former,
that is, we measure the position of the nearby particle. We can then
predict, with 100\% confidence, the exact position of the particle
that we sent off towards Mars.

Alternatively, if we had chosen to measure the momentum of the nearby
particle, we could have predicted, also with 100\% confidence, what a
momentum measurement on the Mars-bound particle would find. We say
that the preparation of the two particles has
\emph{entanglement}. (Schr\"odinger introduced this term in
1935~\cite{Schroedinger:1935}, though Weyl understood the mathematics
by 1931, calling it a variety of ``Gestalt''~\cite{Weyl:1931}.)  The
mathematics of quantum physics implies that entanglement is a kind of
``all-purpose flour''~\cite{Fuchs:2010}.  Given the ability to make an
entangled preparation, we can turn that entanglement into a
correlation between the observables of our choice; in this case, we
can get 100\% correlation between the results of position measurements
\emph{or} between the results of momentum measurements.

It was important for EPR that the second particle be far away, because
then they thought they could safely assume that it cannot be
influenced by what we choose to do on the first particle. As Einstein
said, in the course of developing his thought-experiments on the
topic: ``Thus, without any experiment on [a particle], it is possible
to predict, according to a free choice, \emph{either} the momentum
\emph{or} the position of [it] with in principle arbitrary
accuracy''~\cite{Howard:1990}. Schr\"odinger later used the analogy of
a student getting quizzed in an oral
examination~\cite{Schroedinger:1935}. The student is too tired to
answer more than one question, so they will get the first question
correct, but their answer to the second will be pretty much random.
However, because they do not know ahead of time what the first
question will be, they have to know \emph{all} the answers going in.
The distant particle is like this fatigued student. It has to ``know''
both the position and the momentum of the nearby particle, even though
in any specific experiment it can reveal only one of the two. This
amounts to saying that it must have both a position and a momentum
before we measure it.

The key conceptual ingredients of the EPR argument are as
follows. First, it must be the case that once a particle is far enough
away, it really is free from influence. If the distant particle could
receive instant secret messages from the first, the whole argument
would go out the window. Second, we have to be able to choose one of
two actions to take on the nearby particle, and we have to be able to
use what we deduce about the situation where we pick one option to
make an inference about the other, mutually exclusive situation.
Third is the assumption nowadays called the \emph{EPR criterion of
  reality}:
\begin{quotation}
  \noindent If, without in any way disturbing a system one can [gather
  the information required to] predict with certainty (i.e., with
probability equal to unity) the value of a physical quantity, then
there exists an element of physical reality corresponding to this
physical quantity.
\end{quotation}

Note what EPR do \emph{not} do. They do not measure the position of
one particle and the momentum of the other: That would leave open the
possibility of a disturbance to both particles, thereby making their
criterion of reality inapplicable. Also, they do not measure both
observables in succesion on the same particle. Were one to do that,
the first measurement would change the situation, and the ability to
make a prediction about the distant particle would be lost. This is
also a point that Schr\"odinger emphasized~\cite{Schroedinger:1935}.

A later variation of the EPR thought-experiment, introduced by David
Bohm, has a more ``digital'' feel. Instead of position and momentum,
the observables in Bohm's version have binary
outcomes~\cite{Bohm:1951}. In the Bohm version, two atoms emerge from
a common source and speed in opposite directions, each passing between
the poles of a specially-shaped magnet.  The result the experimenter
observes is whether the atom swerves toward the north pole or toward
the south pole of its magnet.  (Bohm describes this thought-experiment
in terms of the ``spins'' of the two particles. ``Spin'' is one of the
very many words that physics has taken from ordinary language and used
in an esoteric way.) Instead of choosing to measure either the
position or the momentum, the experimenter chooses which
\emph{direction} to align the two magnets.  If both magnets are
aligned along the same axis, then whenever the first atom swerves to
the north pole, the second will swerve to the south.  But if the
second magnet is rotated $90^\circ$ with respect to the first, then
the result of the first measurement is no good at all for predicting
the outcome of the second.  

Let's say that each magnet can either stand vertically or be laid down
flat horizontally.  So, we can specify a choice of ``detector
settings'' by writing one of the four possibilities: HH, VV, VH or HV.
Likewise, we can record the outcomes as NN, SS, SN or NS.  Bohm showed
that with the proper preparation of the two atoms, when the magnet
orientations are HH or VV, the outcomes are always SN or NS, and there
is no bias between these two alternatives.  But if the magnets have
different orientations, HV or VH, then all four outcomes occur with
equal probability.

In Schr\"odinger's analogy, the second atom is like the fatigued
student. It has to ``know'' which direction to swerve no matter whether
the magnet through which it passes is H or V. But following the
textbook rules of quantum physics, the mathematical description of how
the two atoms are prepared --- their ``wave function'' or ``$\psi$
function'' --- does not include this information. Indeed, according to
the ``a quantity doesn't exist unless we measure it'' slogan, quantum
physics seems to have no room for information of that kind.

One can carry the argument a step further. Suppose that you have one
half of a Bohm-style pair of atoms, and you haven't yet decided
whether you want to do the H experiment or the V experiment. \emph{If}
you pick the H experiment, you will be able to make an H prediction,
and so using the EPR criterion, you can say that the other atom has an
H property. Likewise, \emph{if} you pick the V experiment, you will be
able to use the EPR criterion to say that the other atom has a V
property. Before you do a measurement and have the result in hand, you
can't employ the EPR criterion to make a specific statement. For
example, until you do a V measurement on the first atom, you can't say
what the \emph{value} of the second atom's V property is --- whether
it's ``swerve north'' or ``swerve south''. But you \emph{can} say,
ahead of time, that the atom must have a V property, and also an H
property. You don't know \emph{which} value they have, but they've
got to be there. And because they aren't part of how quantum mechanics
describes the situation, quantum mechanics must be incomplete --- or
so EPR would say.

We won't need to worry about this extra wrinkle too much. The problems
with many retellings of EPR can be seen without it. The idea of
deducing that properties exist ahead of time without saying exactly
what values they take really comes into its own in later arguments
that cut into the quantum mysteries more
deeply~\cite{Fuchs:2016}. Other people came to this topic after Bohm,
and we will discuss some of their contributions momentarily. First, we
need to clear the air of misconceptions about what EPR said.

\section{Becker}
\label{sec:becker}
Our first example comes from Adam Becker's recent book,
\booktitle{What is Real?: The Unfinished Quest for the Meaning of
  Quantum Physics} (Basic Books, 2018). I am often looking for books
that might help my relatives understand my job and why I do
it. Unfortunately, despite the vigor of its prose, this one does not
earn a recommendation. The following is what the first through third
printings had to say about the EPR thought-experiment.
\begin{quotation}
  \noindent In quantum physics, the situation is a little
  trickier. According to the Copenhagen interpretation, particles
  don't have properties like position or momentum (or anything else)
  until those properties are measured. But, EPR argued, measurements
  made on one particle couldn't instantly affect another particles
  [sic] far away. So, to get around the uncertainty principle, just
  wait until particles A and B are very far apart, then find the
  momentum of A. Measuring A's momentum lets you infer B's momentum
  without disturbing B at all. Then simply measure the position of
  B. Now you know B's position and momentum, to arbitrary precision,
  at the same time. Therefore, argued EPR, a particle can have a
  definite position and momentum at the same time.
\end{quotation}
The problem here kicks in with the instruction to ``simply measure the
position of B''. This is \emph{not} part of EPR, and for good
reason. EPR were trying to show that a measurable property, like a
particle's position, must exist before it is measured. In order to
make an argument to this effect, they had to introduce a way of
measuring a particle indirectly.  According to Heisenberg, a direct
measurement doesn't just read off a property that already exists.
Thus, when you ``simply measure the position of B'', you are
\emph{not} just revealing a property that B already had.  What EPR
\emph{actually} argued was that in one scenario, when you measure the
momentum of A, you can deduce that B must be carrying a particular
value of momentum; and in another, mutually exclusive scenario, when
you measure the position of A, you can likewise deduce that B must
have a position before you measure it.  By pooling these two
deductions, you then convince yourself that B must have ``a definite
position and momentum'', whether or not you measure either one.
Becker's version \emph{assumes the conclusion of EPR} while attempting
to explain \emph{the argument of EPR} --- a great big logical loop.

There is no choice between alternatives in Becker's description, just
a single procedure to follow. EPR recognized that making a selection
was a significant step, but Bohr took that point further. Much of
Bohr's reply to EPR was devoted to devising a more specific
implementation scheme for their
thought-experiment~\cite{Bohr:1935}. This mattered to Bohr, because he
wished to emphasize that the \emph{choice} of measuring either
position \emph{or} momentum requires selecting one of two mutually
exclusive laboratory setups~\cite[pp.\ 195--6]{Jammer:1974}. We will
return to the topic of Bohr's reply later (\S\ref{sec:interlude}), but
for now, we should underline the fact that missing the ingredient of
choosing between experimental arrangements means that this portrayal
fails to engage with a crucial part of Bohr's thinking.  This is
particularly important because Becker's book is in large part a
polemic against Bohr's viewpoint, or at least a viewpoint that it
attributes to Bohr. Later printings fixed this error, but left others
(we will touch on a systemic problem in \S\ref{sec:interlude}).  I
bring this erratum to the forefront not just as a conversation piece
about how quality control does or does not happen in pop science ---
although that is a conversation I believe we need to have! --- but
also because it provides a clean introduction to the other misreadings
of EPR that we will survey as we go along.

We can step through the problem with the Bohm version of the
thought-experiment.  One of the empirical facts about atoms being
tested in this way, by being fired through these specially-shaped
magnets, is that if an atom swerves to the north after going through
an H magnet, it will swerve to the north again if immediately passed
through a second H magnet.  Likewise, south-swervers in one H test
remain south-swervers in a second H test made just after the first.
The same holds true for one V measurement followed by another V.  But
if we follow an H test with a V, sending into the V test only those
that swerved a certain way (say, to the north), then the results of
the V test will be \emph{random.}  In general, the result of tests in
immediate succession are the \emph{same} if the orientations are the
same (H then H, or V then V); but the output of the second test is
\emph{random} if the orientations are different (H then V, or V then
H).

Let's say we pick the measurement on particle $A$ to be with a magnet
oriented horizontally. Then we can predict with 100\% confidence the
result of a measurement on particle $B$, if the magnet through which
particle $B$ passes is also oriented horizontally.  By the EPR
criterion, we deduce that ``direction of horizontal swerve'' is a
property that particle $B$ is carrying.  Now, we consider what happens
if we instead measure particle $B$ by passing it through a vertical
magnet.  Does the result we get, north or south, tell us anything
about what intrinsic properties particle $B$ might have?  Well, why
should it?  The results of that V test will be random!  Remember, when
the two magnets are chosen to be HV, then the results are NN, SS, SN
and NS with equal probability across the board.  Whatever the result
of the first (H) measurement, the second (V) can register N or S, with
no bias between them.  We \emph{have no grounds for saying} that
particle $B$ had any ``direction of vertical swerve'' property before
we started measuring.  \emph{If} we measure particle $A$ with a
vertical magnet first, \emph{then} EPR would say that particle $B$ has
a ``direction of vertical swerve'' property.  On the \emph{further}
assumption that we can pool the deductions from these two mutually
exclusive scenarios (EPR found this reasonable, Bohr did not), then we
can finally say that particle $B$ must always be carrying properties
that specify how it will swerve in either orientation.

Now, if you have accepted the EPR gospel and convinced yourself that
particles really do carry intrinsic ``direction of horizontal swerve''
and ``direction of vertical swerve'' properties, then you can do the
following.  First, measure particle $A$ with an H magnet, and deduce
the ``direction of horizontal swerve'' property of particle $B$.
Then, measure particle $B$ with a V test.  \emph{Provided that you
  have already concluded} that particle $B$ has both properties, then
you can say that your test with the V magnet has revealed what
particle $B$'s ``direction of vertical swerve'' property was.  So, you
can then say that the source prepared particle $B$ with definite
values of those two properties: ``swerve to N in an H test'' and
``swerve to S in a V test'', for example.  But here's where the
caveats come into play.  First, you can only say this \emph{if} you
have already gone through the whole rigmarole of pooling deductions
from different scenarios.  You need EPR's experiment, not Becker's.
Second, the value you deduced for the property of $B$ that you did not
directly measure --- its ``direction of horizontal swerve'' property,
in this case --- does you no good.  You can't use it to calculate
anything \emph{else,} because the V test you applied directly to
particle $B$ scrambled that property!\footnote{Schr\"odinger ran into
  a related problem~\cite{Schroedinger:1935}. He convinced himself,
  essentially by EPR's logic, that a particle has to have an intrinsic
  value of momentum and, simultaneously, an intrinsic value of
  position.  Then, having granted that much, he argued that we could
  get a number for the momentum by directly measuring it and also get
  a number for the position by inferring it indirectly from an observation
  of the other particle.  But, he found, the two numbers so obtained
  just did not work in equations the way they should. Schr\"odinger found
  the situation he had wedged himself into ``bewildering''; to explore
  his bewilderment more fully would require a higher level of
  mathematics than this essay is striving to maintain.}

\section{Horgan}

Becker's presentation of EPR went awry, but he did present it as
making inferences from position to position and momentum to momentum.
We can find expositors who do worse.
John Horgan's \booktitle{The End of Science} (Addison-Wesley, 1996)
has this to say about the EPR thought-experiment:
\begin{quotation}
  \noindent According to the standard model of quantum mechanics,
  neither particle has a definite position or momentum before it is
  measured; but by measuring the momentum of one particle, the
  physicist instantaneously forces the other particle to assume a
  fixed position---even if it is on the other side of the galaxy.
\end{quotation}
From the momentum of one, we get the position of the other!
Remarkable. This is not EPR.

Horgan's exposition has the additional problem that it fails to
distinguish between EPR and what others brought to the table many
years later. In 1964, John S.\ Bell built upon the work of Bohm and of
EPR to prove a theorem that pierces more deeply than they did into
what makes quantum physics exotic~\cite{Bell:1964}. Others built upon
Bell's work in turn, devising variations on his ideas that brought
them closer to testing in practice~\cite{Clauser:1969} and making more
general versions~\cite{Braunstein:1990}. The experimentalists have
tested Bell's conclusions in increasingly stringent ways and have
confirmed them with ever-higher standards~\cite{BIG:2018}. Horgan
oversimplifies this history to the point of error, reducing it to a
remark that ``a group of French physicists carried out a version of
the EPR experiment'' --- a considerable slight. (Indeed, Horgan never
mentions Bell at all, a surprising lacuna given his lengthy interview
with, and repeated mentions of, David Bohm.) While \booktitle{The End
  of Science} met with a harsh reception upon its original
publication~\cite{Hoffman:1998}, these issues seem not to have been
raised.

\section{Kaiser}

Our next example is from a more recent book.  In David
Kaiser's \booktitle{How the Hippies Saved Physics} (W.\ W.\ Norton \&
Company, 2011), we find the following.
\begin{quotation}
  \noindent The EPR authors described a source, such as a radioactive
  nucleus, that shot out pairs of particles with the same speed but in
  opposite directions.  Call the left-moving particle ``A,'' and the
  right-moving particle ``B.''  A physicist could measure A's position
  at a given moment, and thereby deduce the value of B's
  position. Meanwhile, the physicist could measure B's momentum at
  that same moment, thus capturing knowledge of B's momentum and
  simultaneous position to any desired accuracy.
\end{quotation}
This is the same error that we saw in the Becker example
(\S\ref{sec:becker}), albeit with the roles of position and momentum
exchanged.  Without the ingredient of choice, and the assumption that
we can combine deductions based on mutually exclusive scenarios, we
can't make the argument go.  We could measure A's position and thus
infer a value for B's position, but nothing in this version leads us
to think that the momentum measurement on B has to be revealing a
pre-existing property of B.\footnote{After this article appeared on
  the arXiv, Kaiser wrote to say that he had probably been writing
  with later Bell-test experiments in mind and partly mixed up the
  scenarios, and that he would correct this passage if he ever has the
  opportunity to update the book (e-mail, 15 September 2019).}

\section{Weinberg}
\label{sec:weinberg}
Serious physicists might not care about errors in a popular book.
After all, it is only a question of anyone outside of physics
understanding or valuing what we do. But even a physicist who manages
to maintain that attitude might, reluctantly, acknowledge that
errors in a textbook can be significant.

We turn, therefore, to the 2013 edition of Steven Weinberg's
\booktitle{Lectures on Quantum Mechanics} (Cambridge University
Press). In chapter 12, ``Entanglement'', we find the following.
\begin{quotation}
  \noindent Einstein \emph{et al.}\ imagined that an observer who
  studies particle 1 measures its momentum, and finds a value $\hbar
  k_1$. The momentum of particle 2 is then known to be $-\hbar k_1$,
  up to an arbitrarily small uncertainty. But suppose that the
  observer then measures the position of particle 1, finding a
  position $x_1$, in which case the position of particle 2 would have
  to be $x_1 + x_0$. We understand that the measurement of the
  position of particle 1 can interfere with its momentum, so that
  after the second measurement the momentum of particle 1 no longer
  has a definite value. But how can the second measurement interfere
  with the momentum of particle 2, if the particles are far apart? And
  if it does not, then after both measurements particle 2 must have
  both a definite position and a definite momentum, contradicting the
  fact that these observables do not commute.
\end{quotation}
Here, at least, we are focusing on particle 1 and trying to make
deductions about particle 2. The trouble is that the observer measures
the \emph{same particle} twice, which is not at all equivalent to
making a choice between measuring position \emph{or} momentum.

Weinberg repeates this mistake when he describes Bohm's version of the
EPR experiment, which uses spin degrees of freedom.
\begin{quotation}
  \noindent If the $z$-component of the spin of particle 1 is
  measured, it must have a value $\hbar/2$ or $-\hbar/2$, and then the
  $z$-component of the spin of particle 2 must correspondingly have a
  value $-\hbar/2$ or $+\hbar/2$, respectively. This [is] not
  mysterious --- the particles were once in contact, so it is not
  surprising that the $z$-components of their spins are strongly
  correlated. Following this measurement, suppose that the
  $x$-component of the spin of particle 1 is measured. It will be
  found to have the value $\hbar/2$ or $-\hbar/2$, and the
  $z$-component of particle 1's spin will no longer have a definite
  value. Also, because the system has zero total angular momentum, the
  spin of particle 2 will then have $x$-component $-\hbar/2$ or
  $\hbar/2$, and its $z$-component will not have a definite value.
\end{quotation}
Again, we have the measurements being applied in succession on the
same object.  This explicitly contradicts Bohm, who insists, ``the
investigator can measure either the $x$, $y$, or $z$ component of the
spin of particle No.~1, but not more than one of these components, in
any one experiment''~\cite[p.\ 614]{Bohm:1951}. The final sentence
adds to the confusion, as it appears to say that the second
measurement, the one on the $x$-component, will force the previously
established value of the $z$-component into remission. In addition to
missing the point that the first measurement consumes the
entanglement, this line also confounds the EPR criterion of reality,
which is based on the assumption that the measurement of one particle
\emph{does not disturb} the other.

It is worth noting that Weinberg corrected this error in the second
edition of his book (2015). ``But suppose that the observer then
measures the position of particle 1'' becomes ``But suppose that the
observer instead measures the position of particle 1''; in the later
passage, the description of successive measurements is replaced with
``But the observer could have measured the $x$-component of the spin
of particle 1 instead of its $z$-component''. The final sentence of
the passage (``Also, because the system has\ldots''), which compounded
the problem, was removed altogether.\footnote{I am reliably informed
  that these errors were fixed because N.\ D.\ Mermin pointed them out
  to Weinberg.}

This brings us to an interesting historical point. When reading older
books, one often gets the impression that quantum mechanics is a
limitation on our abilities. Nature, we are told, cunningly frustrates
our right to know simultaneous values for noncommuting observables to
arbitrary precision. A different perspective gained prominence in the
1990s: Instead of frustrating us, quantum physics is a source of new
opportunities~\cite{Caves:2013}. Puzzling phenomena, like interference
of probability amplitudes or quantum entanglement, became seen as
\emph{resources} that might enable new feats of communication and
computation. This change in perspective brings with it a new
intuition. Resources get consumed! A natural question from this
perspective is, then, how long does it take to use up a given quantity
of entanglement?  The Bohm version of EPR illustrates a fundamental
part of the answer: We get one prediction, but not two.
  
\section{Pagels}
\label{sec:pagels}
Next, we return to the more popular literature, but to an instance of
it written by a physicist: \booktitle{The Cosmic Code: Quantum Physics
  as the Language of Nature} (Bantam, 1982), by Heinz R.\ Pagels.  A
review by Mermin pointed out that Pagels' presentation of the EPR
experiment ``gets the central point entirely
wrong''~\cite{Mermin:1983}.  Not having here the space constraints of
a single-column book review, we can quote the relevant passage in
full:
\begin{quotation}
  \noindent Two particles, call them 1 and 2, are sitting near each
  other with their positions from some common point given by $q_1$ and
  $q_2$ respectively. We assume the particles are moving and that
  their momenta are $p_1$ and $p_2$. Although the Heisenberg
  uncertainty relation implies that we cannot simultaneously measure
  $p_1$ and $q_1$ or $p_2$ and $q_2$ without uncertainty, it does
  allow us to simultaneously measure the \emph{sum} of the momenta $p
  = p_1 + p_2$ and the \emph{distance between} the two particles $q =
  q_1 - q_2$ without difficulty. The two particles interact, and then
  particle 2 flies off to London while 1 remains in New York. These
  two locations are so far apart that it seems reasonable to suppose
  that what we do to particle 1 in New York should in no way influence
  particle 2 in London---the principle of local causality. Since we
  know that the total momentum is conserved---it is the same before
  the interaction as after---if we measure the momentum $p_1$ of the
  particle in New York, then by subtracting this quantity from the
  known total momentum $p$, we deduce exactly the momentum $p_2 = p -
  p_1$ of particle 2 in London. Likewise by next exactly measuring the
  position $q_1$ of the particle in New York we can deduce the
  position of particle 2 in London by subtracting the known distance
  between the particles, $q_2 = q_1 - q$. Measuring the position of
  the New York particle will disturb our previous measurement of its
  momentum $p$, but it should not (if we believe in local causality)
  alter the momentum $p_2$ we just deduced for the particle far away
  in London. Hence we have deduced both the momentum $p_2$ and the
  position $q_2$ of the particle in London without any uncertainty.
\end{quotation}
How exactly did we do that again? ``By next exactly measuring'' --- it
is the same problem we saw in the Weinberg example above. (Also, in
order to find a numerical value for the difference $q_1 - q$, we have
to know a number for $q$; but that is a value we get from making a
joint measurement on both particles, so we cannot say that the London
particle is undisturbed.)

Shortly thereafter, Pagels addresses a related topic, von Neumann's attempt
to prove that no hidden-variable completion of quantum physics was
possible.
\begin{quotation}
  \noindent Von Neumann's proof was logically flawless, but as Bell
first pointed out, one of the assumptions that went into Von [sic] Neumann's
proof did not apply to quantum theory and therefore the proof was
irrelevant.
\end{quotation}
The assumptions of von Neumann's argument are not assumptions about
quantum theory, but about a conjectural hidden-variable completion of
it, so the claim that one ``did not apply to quantum theory'' is
pretty garbled. Moreover, historically, Bell was not the first to
locate the crucial flaw in von Neumann's argument. Grete Hermann did
so as early as 1935~\cite{Hermann:1935, Hermann:1935b,
  Mermin:2018}. (This fact was overlooked for many years, before Max
Jammer noted it in his book \booktitle{The Philosophy of Quantum
  Mechanics}. Jammer's book came out in 1974, so Pagels, writing not
quite a decade later, could in principle have known the fuller
history, but it's still a rather arcane tale today. A significant
complicating factor is that Hermann herself did not believe in hidden
variables.)  Hermann charged that von Neumann had essentially assumed
his conclusion, making his argument an exercise in circular logic; the
pleasing symmetry of a circle is, in this case, a bad way to be
``flawless''.\footnote{In 1938, Bohr declared, ``[T]he completeness
and self-consistency of the whole formalism is most clearly exhibited
by the elegant axiomatic exposition of von Neumann, which in
particular makes it evident that the fundamental superposition
principle of quantum mechanics logically excludes the possibility of
avoiding the non-causal feature of the formalism by any conceivable
introduction of additional variables''~\cite{Bohr:1938}. It's hard to
say whether he was impressed by the specific theorem or by the
elegance of von Neumann's work overall. A good framework can be more
compelling than a single proof, though it is easier to point to the
latter than to the former. By contrast, at around the same time,
Einstein discussed the theorem with his assistants at
Princeton. According to one of those assistants, Einstein pulled von
Neumann's textbook off his shelf, pointed to the same assumption
critiqued by Hermann, and asked ``\emph{why we should believe in
that}''~\cite{Wick:1995}. I have elsewhere dug more deeply into von
Neumann arcana~\cite{Stacey:2016b}.}

\section{Isaacson}
\label{sec:isaacson}
I'll admit that I had high hopes for Walter Isaacson's
\booktitle{Einstein: His Life and Universe} (Simon \& Schuster, 2007),
which has been widely praised. Here is how Isaacson treats EPR:
\begin{quotation}
  \noindent We can take measurements on the first particle, the
  authors asserted, and from that gain knowledge about the second
  particle ``without in any way disturbing the second particle.'' By
  measuring the position of the first particle, we can determine
  precisely the position of the second particle. And we can do the
  same for the momentum. ``In accordance with our criterion for
  reality, in the first case we must consider the quantity P as being
  an element of reality, in the second case the quantity Q is an
  element of reality.''
\end{quotation}
This is borderline.  A reader who knows about how quantum entanglement
works could read ``And we can do the same for the momentum'' as ``we
can \emph{alternatively} do the same for the momentum'', particularly
when that sentence is taken with the ``first case''/``second case''
language that follows.  (Position and momentum should have been
mentioned in opposite order, but that's a detail I'm willing to pass
over.)  We move in a bad direction when Isaacson turns to Bohr's
reply:
\begin{quotation}
  \noindent [T]he EPR paper did not, as Bohr noted, truly dispel the
  uncertainty principle, which says that it is not possible to know
  \emph{both} the precise position and momentum of a particle \emph{at
    the same moment.} Einstein is correct, that if we measure the
  \emph{position} of particle A, we can indeed know the
  \emph{position} of its distant twin B. Likewise, if we measure the
  \emph{momentum} of A, we can know the \emph{momentum} of B. However,
  even if we can \emph{imagine} measuring the position and then the
  momentum of particle A, and thus ascribe a ``reality'' to those
  attributes in particle B, we cannot \emph{in fact} measure
  \emph{both} these attributes precisely at any one time for particle
  A, and thus we cannot know them both precisely for particle B.
\end{quotation}
This is further over the borderline.  Recall our discussion of
Weinberg (\S\ref{sec:weinberg}): We can certainly imagine
``measuring the position and then the momentum of particle A'', but in
that scenario, we can only make a position prediction for B.  EPR do
not ascribe a reality to both attributes of B because we could measure
both observables in succession on A, but because we could measure
\emph{either} observable on A, and B has no way of knowing which of
the two we chose.

\section{Cassidy}

David C.\ Cassidy's \booktitle{Uncertainty: The Life and Science of
  Werner Heisenberg} (W.\ H.\ Freeman \& Company, 1992) won a Science
Writing Award from the American Institute of Physics~\cite{AIP} and
appears generally well-regarded.  Turning to its treatment of EPR, we
find the following.
\begin{quotation}
  \noindent Their argument entailed, as usual, a thought experiment:
  two independent particles $A$ and $B$ interact for a finite time,
  then separate without further interaction. There is no
  quantum-mechanical reason that one cannot measure the momentum and
  position of $A$ at two different times without disturbing $B$. If
  this is the case, then one should be able to predict with absolute
  certainty the simultaneous momentum and position of $B$ at any time
  after the interaction. Definite values of the momentum and position
  of $B$ at a given instant must therefore be elements of reality.
\end{quotation}
By now, the error is familiar.  After the first measurement, the
entanglement is broken. The resource is consumed. It is
ex-entanglement.

In the examples we have seen so far, measurement procedures that are
mutually exclusive have been blended in ways that obscure the logic of
the EPR thought-experiment.  Weinberg, Pagels, Isaacson and Cassidy
turn the choice between two observables into a story about measuring
both observables in succession.  Becker and Kaiser make a distinct but
related mistake, forcing two observables together in a different way.
Yet another error would be to avoid the issue of choosing between
observables by including only one observable in the story.  In the
next section, we will examine cases of this type.

\section{\booktitle{New Scientist} and Carroll}

The mathematics of entanglement can be quite subtle, but it is
possible to get the concept wrong in a blunt way. One method for doing
so is to ignore, or gloss over, how entanglement differs from
ordinary, classical correlation.  An exchange in
\booktitle{New Scientist} magazine brings this point to the forefront
and furnishes our next example.  A reader wrote in with a question
about quantum entanglement, proposing a possible everyday analogy and
asking whether it was a good one.  The analogy involved two brothers,
Robbie and Fred. One of them carries a green wallet, the other a red
one. They leave home and go in separate directions, until a mugger
reveals the color of Robbie's wallet. At that moment, we know the
color of Fred's wallet as well~\cite{NS:2007}.

There is nothing wrong with a reader of a pop-science magazine
thinking up this analogy and wanting to know if it is oversimplified.
Indeed, we should encourage that kind of curiosity! But there is a
great deal wrong when the magazine prints their letter and says, ``No,
it's exactly right''~\cite{Bacon:2007}.  At the very least, to make
the analogy work at all, we have to be able to measure each of the two
systems in two different ways.  Without that, we can't express the
argument of EPR, nor can we describe Bohm's digital variant, and we
have no shot at exploring the even deeper discoveries that came
after. In short, we are unequipped to draw a meaningful line between
classical and quantum.

A related example comes from Sean Carroll's book \booktitle{From
  Eternity To Here: The Quest for the Ultimate Theory of Time}
(Dutton, 2010).  This book contains a fairly lengthy section on
entanglement and, nominally, the EPR thought-experiment.  Carroll
gives the EPR paper itself a brief mention and credit for introducing
the concept of an entangled $\psi$ function, but he does not address
their motivation, how they wanted to show that quantum particles have
definite values of position and momentum even before those quantities
are measured.  The bulk of the section is a tale of two animals for
whom a joint $\psi$ function is written. When ``Miss Kitty'' is
observed, she is found to be either on the table or on the sofa; when
``Mr.\ Dog'' is observed, he is found to be either in the living room
or in the yard.
\begin{quotation}
  \noindent Even though we have no idea where Mr.\ Dog is going to be
  before we look, if we first choose to look for Miss Kitty, once that
  observation is complete we know exactly where Mr.\ Dog is going to
  be, even without looking for him! That's the magic of entanglement.
\end{quotation}
No, it isn't. It's an unremarkable possibility that could occur in
everyday life.  The entire buildup to this declaration is beside the
point.\footnote{Part of that buildup includes an assertion that a
$\psi$ function must be ``real'' and ``not just a bookkeeping device
to keep track of probabilities'' (p.\ 237). This claim is unfounded;
Carroll's brief argument for it neglects the fact that probabilities
for different, mutually exclusive scenarios do not have to fit
together in the way that classical physics assumes they
do~\cite{Wootters:1986, Appleby:2017b}. This point was appreciated by
Feynman in 1948~\cite{Feynman:1948}, and Born and Heisenberg seem to
have had at least a finger on it as early as 1927~\cite{Born:1928}.}
None of the conceptual or mathematical apparatus of quantum theory is
necessary for Carroll's scenario, and a big sign of why is that the
story considers only one observable, the location, of each character.
\section{Popper}
\label{sec:popper}
Karl Popper is on the short list of \emph{the} philosophers of the
twentieth century. He discussed the EPR thought-experiment at
considerable length in his \booktitle{The Logic of Scientific
  Discovery} (Basic Books, 1959).  In Appendix $^\star$xi, we find
something interesting.  In order to bullet-proof the EPR argument
against Bohr's counter-argument (or what he perceives Bohr's logic to
be), Popper introduces a variation of EPR.
\begin{quotation}
  \noindent [T]he ideas of Einstein, Podolsky, and Rosen allow us, by
  a slight extension of their experiment, to determine simultaneously
  positions \emph{and} momenta of both $A$ and $B$---although the
  result of this determination will have \emph{predictive}
  significance only for the position of the one particle and the
  momentum of the other. For [\,\ldots\!]\ we may measure the position of
  $B$, and somebody far away may measure the momentum of $A$
  accidentally at the same instant, or at any rate before any smearing
  effect of our measurement of $B$ could possibly reach $A$.
\end{quotation}
This is an interesting complication.  As an exercise, let's restate it
in terms of the Bohm version.  The experimenter measures $B$ with one
magnet orientation, let us say the vertical.  Meanwhile, ``at the same
instant'', $A$ is being measured with a horizontal magnet.  The first
problem is that if $A$ interacts with \emph{some other physical body}
first, then the entanglement of the $\psi$ function for $A$ and $B$
will be spoiled.

There is also a problem with the phrase ``at the same instant''.
Remember, in special relativity, simultaneity is relative. Whose
reference frame are we talking about, and why is theirs the only one
that matters?

The next appendix in Popper's book, number $^\star$xii, reproduces a
letter that he received from Einstein.  Einstein's correspondence with
Popper is, to my knowledge, the only time that he himself (absent of
Podolsky and Rosen) used the EPR thought-experiment
specifically. Elsewhere, Einstein preferred a related but distinctly
different argument. We can hardly do better than to quote Einstein
himself here --- the thing about Einstein is that when he's on, he's
\emph{on.} In a 1936 essay~\cite{Einstein:1950}, Einstein wrote the
following.
\begin{quotation}
  \noindent Consider a mechanical system constituted of two partial
  systems $A$ and $B$ which have interaction with each other only
  during limited time. Let the $\psi$ function before their
  interaction be given. Then the Schr\"odinger equation will furnish
  the $\psi$ function after their interaction has taken place. Let us
  now determine the physical condition of the partial system $A$ as
  completely as possible by measurements. Then the quantum mechanics
  allows us to determine the $\psi$ function of the partial system $B$
  from the measurements made, and from the $\psi$ function of the
  total system. This determination, however, gives a result which
  depends upon \emph{which} of the determining magnitudes specifying
  the condition of $A$ has been measured (for instance coordinates
  \emph{or} momenta). Since there can be only \emph{one} physical
  condition of $B$ after the interaction and which can reasonably not
  be considered as dependent on the particular measurement we perform
  on the system $A$ separated from $B$ it may be concluded that the
  $\psi$ function is not unambiguously coordinated with the physical
  condition. This coordination of several $\psi$ functions with the
  same physical condition of system $B$ shows again that the $\psi$
  function cannot be interpreted as a (complete) description of a
  physical condition of a unit system.
\end{quotation}
Note that the EPR criterion of reality doesn't enter into Einstein's
argument --- suggesting, perhaps, that Einstein himself grew
suspicious of it.\footnote{It is possible that Einstein came to
  believe that particle position and momentum could not be among the
  actual properties that quantum systems possess~\cite{Fine}.} What
matters is that we have our choice of measurements on $A$, and that
because $B$ is far away we cannot affect its actual physical mode of
being by what we choose to do with $A$. We can ``steer'' (as
Schr\"odinger called it) the $\psi$ function of $B$ with our choice of
action upon $A$, but we cannot affect the ``physical condition'' of
$B$. Therefore, the $\psi$ function of $B$ is not its physical
condition, only a partial description thereof.

Popper devotes a considerable amount of prose to taking apart Bohr's
reply to EPR, which he finds ``unacceptable'' and in a key place
``\emph{ad hoc}''.  I don't think Popper's critiques really connect,
on the whole, but Popper is hardly to be blamed. Identifying the core
principles of Bohr's thinking, the part where a critique should aim to
land, is extraordinarily difficult to do from his reply to EPR alone.

\section{Interlude: The Shorter Bohr}
\label{sec:interlude}
Bohr has a reputation for being an obscure writer. On the whole, this
reputation is amply justified. His sentences can be exercises for the
topologist. It is perhaps unsurprising that one of the most famous
``Bohr quotes'' was actually said by someone else, as few remarks by
the man himself were so snappy~\cite{Mermin:2004}.

Bohr's reply to EPR is, I suspect, a significant contributor to his
air of obscurity. Consider, for example, the experience of
Fuchs~\cite{Fuchs:2014} when he spoke on the topic at a conference:
\begin{quotation}
  \noindent My original title for the talk had been ``Why I Never
  Understood Bohr's Reply to EPR, But Still Liked It''---but I wrote
  it on two overlapping transparencies so that, at the appropriate
  moment, I could strip off the part that said ``But Still Liked It.''
  (I hadn't originally intended to do that, but it was the only thing
  I could do with honesty after rereading Bohr.)
\end{quotation}
We should also note that in a collection of classic historical papers
on quantum physics edited by Wheeler and Zurek~\cite{Wheeler:1983},
two pages of Bohr's reply to EPR were reproduced in the wrong order,
and the error made it all the way to print (pages 148 and 149 in the
book should be reversed).

In my view, Bohr was significantly more clear a few years later, at
the Warsaw conference in 1938~\cite{Bohr:1938}. Using his Warsaw
lecture as a guide, I believe it is possible to streamline Bohr's
reply to EPR rather significantly.

EPR write, near the end of their paper, ``[O]ne would not arrive at
our conclusion if one insisted that two or more physical quantities
can be regarded as simultaneous elements of reality \emph{only when
  they can be simultaneously measured or predicted.}''

The response that Bohr could have made: ``Yes.''

EPR briefly consider the implications of this idea and then dismiss it
with the remark, ``No reasonable definition of reality could be
expected to permit this.'' 

But that is exactly what Bohr did.  A possible reply in the Bohrian
vein: ``Could a `reasonable definition of reality' permit so basic a
fact as the simultaneity of two events to be dependent on the
observer's frame of reference? Many notions familiar from everyday
life only become well-defined in relativity theory once we fix a
Lorentz frame. Likewise, many statements in quantum theory only become
well-defined once we have given a complete description of the
experimental apparatus and its arrangement. If you set up your
laboratory to do a position measurement, you can make a position
prediction. If you instead arrange your equipment to do a momentum
measurement, you can make a momentum prediction. And before you take
your equipment off the shelf, when all your devices are still in the
cabinet, your laboratory is in the null configuration. Trying to blend
deductions from multiple different, mutually exclusive arrangements
will give a meaningless result, just like mixing up reference frames
in relativity will make you blunder into what looks like a paradox.''

This is not a quote from anywhere in Bohr's writings, but it is fairly
in the tradition of his Warsaw lecture, where he put considerable
emphasis on what he felt to be ``deepgoing analogies'' between quantum
theory and relativity.
\begin{quotation}
  \noindent In spite of all differences in the physical problems
  concerned, relativity theory and quantum theory possess striking
  similarities in a purely logical aspect. In both cases we are
  confronted with novel aspects of the observational problem,
  involving a revision of customary ideas of physical reality, and
  originating in the recognition of general laws of nature which do
  not directly affect practical experience. The impossibility of an
  unambiguous separation between space and time without reference to
  the observer, and the impossibility of a sharp separation between
  the behavior of objects and their interaction with the means of
  observation are, in fact, straightforward consequences of the
  existence of a maximum velocity of propagation of all actions and of
  a minimum quantity of any action, respectively.
\end{quotation}
Here, Bohr slides neatly from a fairly everyday definition of
``action'' to a more technically inclined one: Planck's constant is a
quantum of ``action'' in the sense codified by Lagrangian
mechanics.\footnote{Bohr has the reputation of tumbling into
  solipsism, but the Warsaw lecture puts a dart in the neck of
  that. ``In the first place,'' he insists, ``we must recognize that a
  measurement can mean nothing else than the unambiguous comparison of
  some property of the object under investigation with a corresponding
  property of another system, serving as a measuring instrument''. The
  twist is that such properties are not meaningful without ``taking
  the whole experimental arrangement into consideration''. (Becker's
  book, in all its printings, is among those that make much of the
  supposed anti-realism of Bohr and the ``Copenhagen interpretation''.)
  There is also an intriguing bit of symmetry in Bohr's
  considerations: ``defining the initial state'' and specifying the
  observable properties one is trying to predict are \emph{both} part
  of ``the fixation of the external conditions'', and it is only the
  combination of conditions ``of both kinds which constitutes a
  well-defined phenomenon''. For Bohr, there is simply no room for the
  ``collapse of the wavefunction'' to be problematic: We admit the
  need for it at the same time we write a wavefunction in the first
  place.}

I think this analogy is worth chasing down.  But to do so seriously,
we have to ask a deepgoing question: \emph{Can a classical theory have
  a minimum quantity of action?}  In other words, does the essence of
the quantum really live in~$\hbar$?  Answering this question will bring us
back to EPR, and then take us beyond.

\section{EPR and Beyond the Intrinsic}
\label{sec:beyond}
Suppose you had a fairly solid grasp of classical mechanics, but no
quantum physics.  One day, you hear about Heisenberg's uncertainty
principle, and you are told in solemn terms that it cannot be beat.
How might you incorporate this into your understanding of physics?

An entirely sensible way of going about it would be to say, ``All
right, \emph{exactly} knowing the position or the momentum of
\emph{anything} was an idealization that I never achieved in practice,
nor did I ever need to.  Really, there was always fuzz in my knowledge
of whereabouts in phase space any system might be. What your
`uncertainty principle' means is that something will stop me from
making my \emph{probability distributions} on phase space arbitrarily
narrow.  Your constant $\hbar$ sets the scale beyond which further
precision is impossible.''

We can formulate a theory on these lines. Remarkably, when we take
care to develop it consistently, the theory which results is
\emph{operationally equivalent} to a \emph{subtheory} of quantum
physics.  Specifically, this procedure reproduces what is technically
known as \emph{Gaussian quantum physics,} the portion of
nonrelativistic quantum mechanics in which all the ``Wigner
representations'' of states and processes never use negative
numbers~\cite{Bartlett:2012}. Though this is only a thin slice of the
possibilities of full-fledged quantum mechanics, the number of
phenomena that the Gaussian subtheory supports is remarkable. These
phenomena include complementary and noncommuting measurements,
teleportation, key distribution, channel-state duality, the
purifiability of mixed states and more.  By construction, all of these
arise in a theory which admits a natural completion just how Einstein
would have wanted: Each system has its own, intrinsic ``physical
condition'', and the $\psi$ function of a system merely expresses what
the experimentalist \emph{knows} about that system.

This restricted theory, with its natural narrative of underlying
classical variables, includes EPR pairs. The way it does so follows
directly from the original presentation by EPR themselves. In this
theory, an EPR pair is a set of two particles prepared in such a way
that they have perfect correlation between their positions and perfect
anti-correlation between their momenta. The observer is ignorant of
the position and the momentum of either particle, but she knows
everything about how their values are related.

We can equally well invent a theory of this type for the ``digital''
version of EPR that Bohm introduced.  In this theory, each atom has a
``physical condition'' that can take one of four possibilities, but we
can never narrow down our knowledge about that physical condition more
than halfway.  That is, we can never have more than \emph{one} bit of
information about a physical condition that takes \emph{two} bits to
describe fully.  With this restriction, we can know everything about
how two atoms are correlated, without knowing anything about the
physical condition of either atom alone~\cite{Spekkens:2007}.

Thus, there is nothing \emph{fundamentally} nonclassical about the
notion of an EPR pair itself, or about the phenomenon of entanglement
--- or, \emph{pace} Bohr, about the existence of a minimum quantum of
action.  To really dig into the enigma of quantum theory, we have to
find something which the ``toy theories'' we have described just now
cannot emulate.  Which quantum phenomena can be mocked up in these
theories, and which cannot?  This is the territory of theorems by
Bell, Kochen, Specker and others.  When these theorems make use of
entanglement, they push the idea harder, not contenting themselves
with the mere existence of it~\cite{Bell:1964, Bell:1966, Kochen:1967,
  Mermin:1993, Peres:1995, Conway:2006, Bengtsson:2010}. And, in the
modern perspective where quantum strangeness is a \emph{resource,}
this pertains directly to the question of how to make quantum
computers go~\cite{Gottesman:1999, Veitch:2014}.

One way to make this way of thinking more precise is to ask how
exactly quantum theory resists being fit into the phase-space
picture. In classical mechanics, if we know the exact coordinates of a
system in phase space, we can predict anything else.  Perhaps
surprisingly, quantum theory allows for a relaxed version of this:
instead of a full-fledged phase space, an experiment with the property
that, if one has statistics for the potential outcomes of it, one can
calculate the probabilities for the possible outcomes of any
\emph{other} measurement~\cite{Prugovecki:1977, Busch:1991,
  Dariano:2004}.  Another way to express this is that, if we are
sufficiently clever, we can establish a standard \emph{reference
measurement.} Having probabilities for all things that could happen
when we apply the reference measurement to a system is mathematically
the same as writing a quantum state for that system. Instead of
talking about ``$\psi$ functions'' like in the old days, all we need
are those probabilities. In the Bohm version of EPR, our systems of
interest are ``spins''; if a particle is ``spin-$j$'', then a
reference measurement for that spin degree of freedom must have at
least $(2j+1)^2$ outcomes~\cite{Peres:1998, Hardy:2001,
  Medendorp:2011}.  The degree to which such a reference measurement
can be made to resemble the classical ideal of reading off the
phase-space coordinates is a measure of how far quantum theory
deviates from classical physics~\cite{Fuchs:2002, DeBrota:2018,
  Stacey:2018}. Getting a clean and precise expression of this
deviation requires finding the \emph{optimal} reference
measurements. To the surprise of the people who have worked on this
topic, the question of finding the optimal reference measurements
turns out to have intriguing connections with far-flung areas of
mathematics~\cite{Appleby:2016, Stacey:2016, Appleby:2017,
  Stacey:2017, Fuchs:2017b, Kopp:2018}.

Another line of attack uses that idea of knowing, even before you pick
which measurement to do, that you can invoke the EPR criterion. Recall
how the argument went: If you do the H measurement on one half of an
EPR pair, you'll be able to predict the result of an H measurement
upon the other half. Likewise, if you do the V measurement on the
first particle, you'll then be able to predict the result of a V
measurement upon the second particle. So, before you set up your
magnet in either orientation, you know that you \emph{will be able} to
make \emph{some} prediction. You can tell ahead of time that you will
be able to employ the EPR criterion and say that the second particle
has \emph{some} definite value of \emph{a} property. You just don't
yet know which value it will turn out to be. On the assumption that
the second particle really is undisturbed (and also that the EPR
criterion is good), you can already say that the second particle must
have a V value and an H value.

Several people noticed that if you push this argument harder, it
breaks down. If we consider not just \emph{two} alternatives, like H
versus V, but instead a well-chosen menu of possible experiments with more
options, then the set of properties deduced for the second particle
will be \emph{logically inconsistent with itself.} Chasing through the
equations, one finds that in some measurement, the particle is
guaranteed to swerve north, and also guaranteed to swerve south. This
inconsistency forces us back to our assumptions: Something we have
presumed must be wrong~\cite{Fuchs:2016}. 

If you ask me, Einstein was dead on the money when he said that a
$\psi$ function is not a state of being. But the work that has come
since, like the study of what can and cannot arise in a toy theory,
tells us that a $\psi$ function also is not just knowledge about an
intrinsic ``physical condition''. Bohr was right not to think that the
next chapter would reveal a classical layer underlying the quantum.
Making progress and clearing up the metaphysical mess left behind by
the founders requires taking a radical position --- though, perhaps,
the most radical step is arguing that a physicist should \emph{care}
about philosophical matters.  For starters, we can go back to EPR and
take one more look at their criterion of reality.  They speak of
making a prediction ``with probability equal to unity''. Ultimately,
if we want these words to have meaning, we have to decide what means
\emph{probability}. And once we admit the need to find a viable story
for \emph{that,} everything starts to change~\cite{Mermin:2017,
  Fuchs:2017}.

\section*{Acknowledgments}

I thank Chris Fuchs for telling me about the example from Becker's
book, and for suggesting a few other places I should look. Also, I
thank N.\ David Mermin for bringing the Pagels example to my
attention, and for helpful discussions (about the Pagels and Weinberg
passages in particular). Arthur Fine's comments about v3 of this essay
led me to post a v4.


\begin{thebibliography}{999}

\bibitem{Stacey:2015} B.\ C.\ Stacey, \booktitle{Multiscale
  Structure in Eco-Evolutionary Dynamics.} PhD thesis, Brandeis
  University, 2015. \arxiv{1509.02958}.

\bibitem{Einstein:1935} A.\ Einstein, B.\ Podolsky and N.\ Rosen,
  ``\hrefdoi{10.1103/PhysRev.47.777}{Can quantum-mechanical
  description of physical reality be considered
  complete?}''\ \booktitle{Physical Review} \textbf{47} (1935),
  777--80.

\bibitem{Camilleri:2015} K.\ Camilleri and M.\ Schlosshauer,
  ``\hrefdoi{10.1016/j.shpsb.2015.01.005}{Niels Bohr as philosopher of
  experiment: Does decoherence theory challenge Bohr's doctrine of
  classical concepts?}''\\ \booktitle{Studies in History and Philosophy
  of Modern Physics} \textbf{49} (2015), 73--83, \arxiv{1502.06547}.

\bibitem{Howard:1990} D.\ Howard,
  ``\hrefdoi{10.1007/978-1-4684-8771-8_6}{\,`Nicht Sein Kann was Nicht
  Sein Darf,' or the Prehistory of EPR, 1909--1935:\ Einstein's Early
  Worries about the Quantum Mechanics of Composite Systems}.''  In
  \booktitle{Sixty-Two Years of Uncertainty,} edited by
  A.\ I.\ Miller. (Springer, 1990.)
  
\bibitem{Schroedinger:1935} E.\ Schr\"odinger,
  ``\hrefdoi{10.1017/S0305004100013554}{Discussion of probability
  relations between separated systems},'' \booktitle{Mathematical
  Proceedings of the Cambridge Philosophical Society} \textbf{31}
  (1935), 555--63.

\bibitem{Weyl:1931} H.\ Weyl, \booktitle{Gruppentheorie und
  Quantenmechanik} (Methuen, 1931). Translated by H.\ P.\ Robertson as
  \booktitle{The Theory of Groups and Quantum Mechanics} (Dover,
  1950).
  
\bibitem{Fuchs:2010} C.\ A.\ Fuchs, \booktitle{Coming of Age with
  Quantum Information:\ Notes on a Paulian Idea} (Cambridge University
  Press, 2010).

\bibitem{Bohm:1951} D.\ Bohm, \booktitle{Quantum Theory}
  (Prentice-Hall, 1951).

\bibitem{Fuchs:2016} C.\ A.\ Fuchs and B.\ C.\ Stacey, ``Some Negative
  Remarks on Operational Approaches to Quantum Theory.''  In
  \booktitle{Quantum Theory: Informational Foundations and Foils,}
  edited by G.\ Chiribella and R.\ W.\ Spekkens. (Springer, 2016.)
  \arxiv{1401.7254}.

\bibitem{Bohr:1935} N.\ Bohr, ``\hrefdoi{10.1103/PhysRev.48.696}{Can
  quantum-mechanical description of reality be considered
  complete?}''\ \booktitle{Physical Review} \textbf{48} (1935),
  696--702.

\bibitem{Jammer:1974} M.\ Jammer, \booktitle{The Philosophy of Quantum
  Mechanics:\ The Interpretations of Quantum Mechanics in Historical
  Perspective} (Wiley-Interscience, 1974).

\bibitem{Bell:1964} J.\ S.\ Bell, ``On the Einstein Podolsky Rosen
  paradox,'' \booktitle{Physics} \textbf{1} (1964),
  195--200. \url{https://cds.cern.ch/record/111654/files/vol1p195-200_001.pdf}.

\bibitem{Clauser:1969} J.\ F.\ Clauser, M.\ A.\ Horne, A.\ Shimony and
  R.\ Holt, ``\hrefdoi{10.1103/PhysRevLett.23.880}{Proposed experiment
    to test local hidden-variable theories},'' \booktitle{Physical
    Review Letters} \textbf{23} (1969), 880--84.

\bibitem{Braunstein:1990} S.\ L.\ Braunstein and C.\ M.\ Caves,
  ``\hrefdoi{10.1016/0003-4916(90)90339-P}{Wringing out better Bell
  inequalities},'' \booktitle{Annals of Physics} \textbf{202} (1990),
  22--56.

\bibitem{BIG:2018} The BIG Bell Test Collaboration,
  ``\hrefdoi{10.1038/s41586-018-0085-3}{Challenging local realism with
  human choices},'' \booktitle{Nature} \textbf{557} (2018), 212--16,
  \arxiv{1805.04431}.
  
\bibitem{Hoffman:1998} D. Hoffman, ``\booktitle{The End of Science:
  Facing the Limits of Knowledge in the Twilight of the Scientific
  Age} [book review],'' \booktitle{Notices of the AMS} \textbf{45}
  (1998), 260--67.
  \url{http://www.ams.org/notices/199802/bookrev-hoffman.pdf}.
 
\bibitem{Caves:2013} C.\ M.\ Caves, ``Quantum information
  science:\ Emerging no more,''  \arxiv{1302.1864}. In \booktitle{OSA
    Century of Optics} (The Optical Society, 2015).
  
\bibitem{Mermin:1983} N.\ D.\ Mermin, ``\booktitle{The Cosmic Code:
  Quantum Physics as the Language of Nature} by Heinz R.\ Pagels,''
  \booktitle{American Scientist} \textbf{71} (1983), 411,
  \href{https://www.jstor.org/stable/27852152}{\texttt{JSTOR:27852152}}.

\bibitem{Hermann:1935} G.\ Hermann, ``\hrefdoi{10.1007/BF01491142}{Die
  Naturphilosophischen Grundlagen der Quantenmechanik},''
  \booktitle{Die Naturwissenschaften} \textbf{42} (1935), 718--21.

\bibitem{Hermann:1935b} G.\ Hermann, ``Die Naturphilosophischen
  Grundlagen der Quantenmechanik,'' \booktitle{Abhandlungen der
    Fries'schen Schule} \textbf{6} (1935), 69--152.
    
\bibitem{Mermin:2018} N.\ D.\ Mermin and R.\ Schack,
  ``\hrefdoi{10.1007/s10701-018-0197-5}{Homer nodded: von Neumann's
  surprising oversight},'' \booktitle{Foundations of Physics}
  \textbf{48} (2018), 1007--20, \arxiv{1805.10311}.

\bibitem{Bohr:1938} N.\ Bohr, ``The causality problem in atomic
  physics.'' In \booktitle{New Theories in Physics} (International
  Institute of Intellectual Co-operation, 1939).

\bibitem{Wick:1995} D.\ Wick, \booktitle{The Infamous Boundary:\ Seven
  Decades of Heresy in Quantum Physics} (Copernicus, 1995).

\bibitem{Stacey:2016b} B.\ C.\ Stacey,
  ``\hrefdoi{10.1098/rsta.2015.0235}{Von Neumann was not a Quantum
  Bayesian},'' \booktitle{Philosophical Transactions of the Royal
  Society A} \textbf{374} (2016), 20150235, \arxiv{1412.2409}.

\bibitem{AIP}
  \url{https://www.aip.org/aip/awards/science-communication/science-communication-award-scientist/david-c-cassidy}.

\bibitem{NS:2007} ``Superluminal siblings,'' \booktitle{New Scientist}
  letters column (19 September 2007).

\bibitem{Bacon:2007} D.\ Bacon, ``\booktitle{New Scientist} May Be
  New, But About That Science?'' \booktitle{The Quantum
    Pontiff} (28 September 2007),
  \url{http://dabacon.org/pontiff/?p=1628}.

\bibitem{Wootters:1986} W.\ K.\ Wootters,
  ``\hrefdoi{10.1007/BF01882696}{Quantum mechanics without probability
  amplitudes},'' \booktitle{Foundations of Physics} \textbf{16}
  (1986), 391--405.
    
\bibitem{Appleby:2017b} M.\ Appleby, C.\ A.\ Fuchs, B.\ C.\ Stacey and
  H.\ Zhu, ``\hrefdoi{10.1140/epjd/e2017-80024-y}{Introducing the
    Qplex: A novel arena for quantum theory},'' \booktitle{The
    European Physical Journal D} \textbf{71} (2017), 197,
  \arxiv{1612.03234}.

\bibitem{Feynman:1948} R.\ P.\ Feynman,
  ``\hrefdoi{10.1103/RevModPhys.20.367}{Space-time approach to
  non-relativistic quantum mechanics},'' \booktitle{Reviews of Modern
  Physics} \textbf{20} (1948), 367--87.
    
\bibitem{Born:1928} M.\ Born and W.\ Heisenberg, ``Quantenmechanik.''
  In \booktitle{Electrons et photons: rapports et discussions du
    cinqui\`eme conseil de physique tenu a Bruxelles du 24 au 29
    octobre 1927 sous les auspices de l'Institut international de
    physique Solvay,} H.\ A.\ Lorentz, ed.\ (Gauthier-Villars,
  1928). Translated in Bacciagaluppi and Valentini's
  \booktitle{Quantum Theory at the Crossroads: Reconsidering the 1927
    Solvay Conference} (Cambridge University Press, 2009),
  \arxiv{quant-ph/0609184}, pp.\ 407--40.
    

  
\bibitem{Einstein:1950} A.\ Einstein,
  ``\hrefdoi{10.1016/S0016-0032(36)91047-5}{Physics and Reality},''
  \booktitle{Journal of the Franklin Institute} \textbf{221} (1936),
  349--82. Reprinted in A.\ Einstein, \booktitle{Out of My Later
    Years} (Philosophical Library, 1950).

  \bibitem{Fine} A.\ Fine,
    ``\hrefdoi{10.1017/S026988970000137X}{Einstein's interpretations
    of the quantum theory},'' \booktitle{Science in Context}
    \textbf{6} (1993), 257--73.
    
\bibitem{Mermin:2004} N.\ D.\ Mermin,
  ``\hrefdoi{10.1063/1.1688051}{What's wrong with this quantum
  world?}'' \booktitle{Physics Today} \textbf{52} (2004), 10.

\bibitem{Fuchs:2014} C.\ A.\ Fuchs, \booktitle{My Struggles with the
  Block Universe: Selected Correspondence, January 2001 -- May 2011}
  (2014). Edited by B.\ C.\ Stacey, with a foreword by
  M.\ Schlosshauer. \arxiv{1405.2390}.

\bibitem{Wheeler:1983} J.\ A.\ Wheeler and W.\ H.\ Zurek,
  eds. \booktitle{Quantum Theory and Measurement} (Princeton
  University Press, 1983).
    
\bibitem{Bartlett:2012} S.\ D.\ Bartlett, T.\ Rudolph and
  R.\ W.\ Spekkens,
  ``\hrefdoi{10.1103/PhysRevA.86.012103}{Reconstruction of Gaussian
    quantum mechanics from Liouville mechanics with an epistemic
    restriction},'' \booktitle{Physical Review A} \textbf{86} (2012),
  012103, \arxiv{1111.5057}.

\bibitem{Spekkens:2007} R.\ W.\ Spekkens,
  ``\hrefdoi{10.1103/PhysRevA.75.032110}{Evidence for the epistemic
  view of quantum states:~A toy theory},'' \booktitle{Physical Review
  A} \textbf{75} (2007), 032110, \arxiv{quant-ph/0401052}.

\bibitem{Bell:1966} J.\ S.\ Bell,
  ``\hrefdoi{10.1103/RevModPhys.38.447}{On the problem of hidden
  variables in quantum mechanics},'' \booktitle{Reviews of Modern
  Physics} \textbf{38} (1966), 447--52.

\bibitem{Kochen:1967} S.\ Kochen and E.\ P.\ Specker, ``The problem of
  hidden variables in quantum mechanics,'' \booktitle{Journal of
    Mathematics and Mechanics} \textbf{17} (1967), 59--87,
  \href{https://www.jstor.org/stable/24902153}{\texttt{JSTOR:\allowbreak{}24902153}}.
  
\bibitem{Mermin:1993} N.\ D.\ Mermin,
  ``\hrefdoi{10.1103/RevModPhys.65.803}{Hidden variables and the two
  theorems of John Bell},'' \booktitle{Reviews of Modern Physics}
  \textbf{65} (1993), 803--15, \arxiv{1802.10119}.

\bibitem{Peres:1995} A.\ Peres, \booktitle{Quantum Theory: Concepts
  and Methods} (Kluwer, 1995).

\bibitem{Conway:2006} J.\ Conway and S.\ Kochen,
  ``\hrefdoi{10.1007/s10701-006-9068-6}{The Free Will Theorem},''
  \booktitle{Foundations of Physics} \textbf{36} (2006), 1441--73,
  \arxiv{quant-ph/0604079}.

\bibitem{Bengtsson:2010} I.\ Bengtsson, K.\ Blanchfield and
  A.\ Cabello, ``\href{10.1016/j.physleta.2011.12.011}{A
    Kochen--Specker inequality from a SIC},'' \booktitle{Physics
    Letters A} \textbf{376} (2010), 374--76, \arxiv{1109.6514}.
  
\bibitem{Gottesman:1999} D.\ Gottesman, ``The Heisenberg
  representation of quantum computers,'' \arxiv{quant-ph/9807006}. In
  \booktitle{Group22: Proceedings of the XXII International Colloquium
    on Group Theoretical Methods in Physics} (International Press,
  1999).

\bibitem{Veitch:2014} V.\ Veitch, S.\ A.\ H.\ Mousavian,
  D.\ Gottesman and J.\ Emerson,
  ``\hrefdoi{10.1088/1367-2630/16/1/013009}{The resource theory of
    stabilizer computation},'' \booktitle{New Journal of Physics}
  \textbf{16} (2014), 013009, \arxiv{1307.7171}.
  
\bibitem{Prugovecki:1977} E.\ Prugove\v{c}ki,
  ``\hrefdoi{10.1007/BF01807146}{Information-theoretical aspects of
  quantum measurement},'' \booktitle{International Journal of
  Theoretical Physics} \textbf{16} (1977), 321--31.

\bibitem{Busch:1991} P.\ Busch,
  ``\hrefdoi{10.1007/BF00671008}{Informationally complete sets of
  physical quantities},'' \booktitle{International Journal of
  Theoretical Physics} \textbf{30} (1991), 1217--27.
  
\bibitem{Dariano:2004} G.\ M.\ d'Ariano, P.\ Perinotti and
  M.\ F.\ Sacchi,
  ``\hrefdoi{10.1088/1464-4266/6/6/005}{Informationally complete
    measurements and group representation},'' \booktitle{Journal of
    Optics B} \textbf{6} (2004), S487, \arxiv{quant-ph/0310013}.

\bibitem{Peres:1998} A.\ Peres and D.\ Terno,
  ``\hrefdoi{10.1088/0305-4470/31/38/003}{Convex probability domain of
  generalized quantum measurements},'' \booktitle{Journal of Physics
  A} \textbf{31} (1998), L671, \arxiv{quant-ph/9806024}.

\bibitem{Hardy:2001} L.\ Hardy, ``Quantum theory from five reasonable
  axioms,''\\ \arxiv{quant-ph/0101012} (2001).

\bibitem{Medendorp:2011} Z.\ E.\ D.\ Medendorp \emph{et al.,}
  ``\hrefdoi{10.1103/PhysRevA.83.051801}{Experimental characterization
  of qutrits using Symmetric Informationally Complete positive
  operator-valued measures},'' \booktitle{Physical Review A}
  \textbf{83} (2011), 051801, \arxiv{1006.4905}.
  
\bibitem{Fuchs:2002} C.\ A.\ Fuchs, ``Quantum mechanics as quantum
  information (and only a little more),'' \arxiv{quant-ph/0205039}
  (2002).

\bibitem{DeBrota:2018} J.\ B.\ DeBrota, C.\ A.\ Fuchs and
  B.\ C.\ Stacey,
  ``\hrefdoi{10.1103/PhysRevResearch.2.013074}{Symmetric
    informationally complete measurements identify the irreducible
    difference between classical and quantum systems},''
  \booktitle{Physical Review Research} \textbf{2} (2020), 013074,
  \arxiv{1805.08721}.

\bibitem{Stacey:2018} B.\ C.\ Stacey, ``Is the SIC outcome there when
  nobody looks?'' \arxiv{1807.07194} (2018).

\bibitem{Appleby:2016} M.\ Appleby, S.\ Flammia, G.\ McConnell and
  J.\ Yard, ``Generating ray class fields of real quadratic fields via
  complex equiangular lines,'' \arxiv{1604.06098} (2016).

\bibitem{Stacey:2016} B.\ C.\ Stacey, ``Geometric and
  information-theoretic properties of the Hoggar lines,''
  \arxiv{1609.03075} (2016).
    
\bibitem{Appleby:2017} M.\ Appleby, S.\ Flammia, G.\ McConnell and
  J.\ Yard, ``\hrefdoi{10.1007/s10701-017-0090-7}{SICs and algebraic
    number theory},'' \booktitle{Foundations of Physics} \textbf{47}
  (2017), 1042--59, \arxiv{1701.05200}.

\bibitem{Stacey:2017} B.\ C.\ Stacey,
  ``\hrefdoi{10.1007/s10701-017-0087-2}{Sporadic SICs and the normed
  division algebras},'' \booktitle{Foundations of Physics} \textbf{47}
  (2017), 1060--64, \arxiv{1605.01426}.
    
\bibitem{Fuchs:2017b} C.\ A.\ Fuchs, M.\ C.\ Hoang and B.\ C.\ Stacey,
  ``\hrefdoi{10.3390/axioms6030021}{The SIC question: History and
  state of play},'' \booktitle{Axioms} \textbf{6} (2017), 21,
  \arxiv{1703.07901}.

\bibitem{Kopp:2018} G.\ S.\ Kopp,
  ``\hrefdoi{10.1093/imrn/rnz153}{SIC-POVMs and the Stark
  conjectures},'' \booktitle{International Mathematics Research
  Notices} (2019), rnz153, \arxiv{1807.05877}.
    
\bibitem{Mermin:2017} N.\ D.\ Mermin, ``Why QBism is not the
  Copenhagen interpretation and what John Bell might have thought of
  it,'' \arxiv[quant-ph]{1409.2454}.  In \booktitle{Quantum
    [Un]Speakables II} (Springer-Verlag, 2017).

\bibitem{Fuchs:2017} C.\ A.\ Fuchs, ``Notwithstanding Bohr, the
  Reasons for QBism,'' \booktitle{Mind and Matter} \textbf{15} (2017),
  245--300, \arxiv{1705.03483}.

\end{thebibliography}
\end{document}